\documentclass[preprint,showpacs,preprintnumbers,amsmath,amssymb]{revtex4}

\usepackage{graphicx}
\usepackage{dcolumn}
\usepackage{bm}
\usepackage{epsfig}

\begin{document}

\preprint{DESY~06-102\hspace{11.5cm} ISSN 0418-9833}
\preprint{July 2006\hspace{14.9cm}}

\title{Charmed-Hadron Fragmentation Functions from CERN LEP1 Revisited}

\author{Bernd A. Kniehl}
\email{bernd.kniehl@desy.de}
\author{Gustav Kramer}
\email{gustav.kramer@desy.de}
\affiliation{{II.} Institut f\"ur Theoretische Physik, Universit\"at Hamburg,
Luruper Chaussee 149, 22761 Hamburg, Germany}

\date{\today}

\begin{abstract}
In Phys.\ Rev.\ D {\bf58}, 014014 (1998) and {\bf71}, 094013 (2005), we
determined non-perturbative $D^0$, $D^+$, $D^{*+}$, $D_s^+$, and $\Lambda_c^+$
fragmentation functions, both at leading and next-to-leading order in the
$\overline{\mathrm{MS}}$ factorization scheme, by fitting $e^+e^-$ data taken
by the OPAL Collaboration at CERN LEP1.
The starting points for the evolution in the factorization scale $\mu$ were
taken to be $\mu_0=2m_Q$, where $Q=c,b$.
For the reader's convenience, in this Addendum, we repeat this analysis for
$\mu_0=m_Q$, where the flavor thresholds of modern sets of parton density
functions are located.
\end{abstract}

\pacs{13.60.-r, 13.85.Ni, 13.87.Fh, 14.40.Lb}
\maketitle

\section{Introduction}

The OPAL Collaboration presented measurements of the fractional energy spectra
of inclusive $D^{*+}$ \cite{opal}, $D^0$, $D^+$, $D_s^+$, and $\Lambda_c^+$
\cite{opal1} production in $Z$-boson decays based on their entire LEP1 data
sample.
Apart from the full cross sections, they also determined the contributions
arising from $Z\to b\bar b$ decays.
This enabled us, partly in collaboration with Binnewies, to determine
lowest-order (LO) and next-to-leading-order (NLO) sets of fragmentation
functions (FF's) for these charmed ($X_c$) hadrons \cite{bkk,kk}.
We took the charm-quark FF to be of the form proposed by Peterson
{\it et al.}\ \cite{pet} and thus obtained new values of the $\epsilon$
parameter, which are specific for our choice of factorization scheme.

We worked in the QCD-improved parton model implemented in the pure modified
minimal-subtraction ($\overline{\mathrm{MS}}$) renormalization and
factorization scheme with $n_f=5$ massless quark flavors (zero-mass
variable-flavor-number scheme).
This scheme is particularly appropriate if the characteristic energy scale of
the considered production process, {\it i.e.}, the center-of-mass energy
$\sqrt s$ in the case of $e^+e^-$ annihilation and the transverse momentum
$p_T$ of the $X_c$ hadron in other scattering processes, is large compared to
the bottom-quark mass $m_b$.
Owing to the factorization theorem \cite{col}, the FF's defined in this scheme
satisfy two desirable properties:
(i) their scaling violations are ruled by the time-like
Dokshitzer-Gribov-Lipatov-Altarelli-Parisi (DGLAP) \cite{dglap} evolution
equations; and
(ii) they are universal.
Thus, this formalism is predictive and suitable for global data analyses.

We verified that the values of the branching and average momentum fractions of
the various $c,b\to X_c$ transitions evaluated at LO and NLO using our FF's
\cite{bkk,kk} are in reasonable agreement with the corresponding results from
OPAL \cite{opal,opal1} and other experiments~\cite{rest}.

We tested the scaling violations of our $D^0$, $D^+$, $D_s^+$, and
$\Lambda_c^+$ FF's \cite{kk} by comparing the fractional energy spectra of
these hadrons measured in non-resonant $e^+e^-$ annihilation at
$\sqrt s=10.55$~GeV \cite{cleo}, 29~GeV \cite{hrs}, and 34.7 \cite{tasso}
with our LO and NLO predictions to find reasonable agreement.
Since events of $X_c$-hadron production from $X_b$-hadron decay were excluded
from the data samples at $\sqrt s=10.55$~GeV, we obtained a clean test of our
charm-quark FF's.

In Refs.~\cite{bkk,kk}, the starting points $\mu_0$ for the DGLAP evolution in
the factorization scale $\mu$ were taken to be $\mu_0=2m_Q$, where $Q=c,b$.
This choice is phenomenologically motivated by the observation that, in
$e^+e^-$ annhilation, which has been providing the most constraining input for
the determinations of FF's, these values of $\mu_0$ represent the very
production thresholds of the respective flavors.
Unfortunately, this choice is inconsistent with the convention underlying
modern sets of parton density functions (PDF's) \cite{pdf}, which prefer to
place the flavor thresholds at $\mu_0=m_Q$.
For the reader's convenience, in this Addendum to Refs.~\cite{bkk,kk}, we thus
repeat the analysis of that papers for the choice $\mu_0=m_Q$, so as to
provide alternative LO and NLO sets of $X_c$ FF's that can be conveniently
utilized together with those PDF's.
The FF's presented below were already used as input for a NLO study
\cite{kkss} of charmed-meson hadroproduction in $p\bar p$ collisions, which
yielded agreement within errors with data collected by the CDF Collaboration
in run II at the Fermilab Tevatron \cite{cdf}.
We note in passing that, in the case of perturbatively induced FF's, which is
quite different from the case of non-perturbative FF's (involving substantial
intrinsic components) under consideration here, the choice $\mu_0=m_Q$ is more
natural, since, at NLO, it avoids finite matching conditions at the flavor
thresholds \cite{cac}.

\section{Results}

In the following, we concentrate on the most important results of
Refs.~\cite{bkk,kk} that are affected by the shift in $\mu_0$.
These include the fit parameters $N$, $\alpha$, $\beta$, and $\epsilon$
defining the $x$ distributions of the $Q\to X_c$ FF's $D_Q(x,\mu^2)$ at
$\mu=\mu_0$,
\begin{eqnarray}
D_c(x,\mu_0^2)&=&N\frac{x(1-x)^2}{[(1-x)^2+\epsilon x]^2},
\label{eq:cff}\\
D_b(x,\mu_0^2)&=&Nx^{\alpha}(1-x)^{\beta},
\label{eq:bff}
\end{eqnarray}
the $\chi^2$ values per degree of freedom ($\chi^2/\mathrm{d.o.f.}$) achieved
in the fits, and the branching fractions $B_Q(\mu)$ and average momentum
fractions $\langle x\rangle_Q(\mu)$,
\begin{eqnarray}
B_Q(\mu)&=&\int_{x_{\rm cut}}^1dx\,D_Q(x,\mu^2),
\label{eq:bq}\\
\langle x\rangle_Q(\mu)&=&\frac{1}{B_Q(\mu)}\int_{x_{\rm cut}}^1dx\,
xD_Q(x,\mu^2),
\label{eq:xq}
\end{eqnarray}
where $x_{\rm cut}=0.1$, at $\mu=2\mu_0$ and $M_Z$.
In the present analysis, we adopt the up-to-date input information from our
2005 paper~\cite{kk}.

Our new results are presented in Tables~\ref{tab:par}--\ref{tab:xav}.
Comparing Tables~\ref{tab:br} and \ref{tab:xav} with the corresponding tables
in Refs.~\cite{bkk,kk}, we observe that the branching and average momentum
fractions are changed very little by the reduction in $\mu_0$.
For a comparison of these observables with experimental data, we refer to
Refs.~\cite{bkk,kk}.

\begin{table}
\begin{center}
\caption{Fit parameters of the charm- and bottom-quark FF's in
Eqs.~(\ref{eq:cff}) and (\ref{eq:bff}), respectively, for the various $X_c$
hadrons at LO and NLO.
The corresponding starting scales are $\mu_0=m_c=1.5$~GeV and 
$\mu_0=m_b=5$~GeV, respectively.
All other FF's are taken to be zero at $\mu_0=m_c$.}
\label{tab:par}
\begin{tabular}{ccccccc}
\hline\hline
$X_c$ & Order & $Q$ & $N$ & $\alpha$ & $\beta$ & $\epsilon$ \\
\hline
$D^0$         & LO  & $c$ & 0.694   & $\cdots$ & $\cdots$ & 0.101    \\
              &     & $b$ & 81.7    & 1.81     & 4.95     & $\cdots$ \\
              & NLO & $c$ & 0.781   & $\cdots$ & $\cdots$ & 0.119    \\
              &     & $b$ & 100     & 1.85     & 5.48     & $\cdots$ \\
$D^+$         & LO  & $c$ & 0.282   & $\cdots$ & $\cdots$ & 0.104    \\
              &     & $b$ & 52.0    & 2.33     & 5.10     & $\cdots$ \\
              & NLO & $c$ & 0.266   & $\cdots$ & $\cdots$ & 0.108    \\
              &     & $b$ & 60.8    & 2.30     & 5.58     & $\cdots$ \\
$D^{*+}$      & LO  & $c$ & 0.174   & $\cdots$ & $\cdots$ & 0.0554   \\
              &     & $b$ & 69.5    & 2.77     & 4.34     & $\cdots$ \\
              & NLO & $c$ & 0.192   & $\cdots$ & $\cdots$ & 0.0665   \\
              &     & $b$ & 20.8    & 1.89     & 3.73     & $\cdots$ \\
$D_s^+$       & LO  & $c$ & 0.0498  & $\cdots$ & $\cdots$ & 0.0322   \\
              &     & $b$ & 27.5    & 1.94     & 4.28     & $\cdots$ \\
              & NLO & $c$ & 0.0381  & $\cdots$ & $\cdots$ & 0.0269   \\
              &     & $b$ & 27.5    & 1.88     & 4.48     & $\cdots$ \\
$\Lambda_c^+$ & LO  & $c$ & 0.00677 & $\cdots$ & $\cdots$ & 0.00418  \\
              &     & $b$ & 41.2    & 2.02     & 5.92     & $\cdots$ \\
              & NLO & $c$ & 0.00783 & $\cdots$ & $\cdots$ & 0.00550  \\
              &     & $b$ & 34.9    & 1.88     & 6.08     & $\cdots$ \\
\hline\hline
\end{tabular}
\end{center}
\end{table}

\begin{table}
\begin{center}
\caption{$\chi^2/\mathrm{d.o.f.}$ achieved in the LO and NLO fits to the
OPAL \cite{opal,opal1} data on the various $X_c$ hadrons.
In each case, $\chi^2/\mathrm{d.o.f.}$ is calculated for the
$Z\to b\overline{b}$ sample ($b$), the full sample (All), and the combination
of both (Average).}
\label{tab:chi}
\begin{tabular}{ccccc}
\hline\hline
$X_c$ & Order & $b$ & All & Average \\
\hline
$D^0$         & LO  & 1.26  & 0.916 & 1.09  \\
              & NLO & 1.10  & 0.766 & 0.936 \\
$D^+$         & LO  & 0.861 & 0.658 & 0.759 \\
              & NLO & 0.756 & 0.560 & 0.658 \\
$D^{*+}$      & LO  & 1.19  & 1.12  & 1.16  \\
              & NLO & 1.07  & 1.01  & 1.04  \\
$D_s^+$       & LO  & 0.246 & 0.111 & 0.178 \\
              & NLO & 0.290 & 0.112 & 0.201 \\
$\Lambda_c^+$ & LO  & 1.05  & 0.117 & 0.583 \\
              & NLO & 1.05  & 0.112 & 0.579 \\
\hline\hline
\end{tabular}
\end{center}
\end{table}

\begin{table}
\begin{center}
\caption{Branching fractions (in \%) of $Q\to X_c$ for $Q=c,b$ and the various
$X_c$ hadrons evaluated according to Eq.~(\ref{eq:bq}) in LO and NLO at the
respective production thresholds $\mu=2m_Q$ and at the $Z$-boson resonance
$\mu=M_Z$.}
\label{tab:br}
\begin{tabular}{cccccc}
\hline\hline
$X_c$ & Order & $B_c(2m_c)$ & $B_c(M_Z)$ & $B_b(2m_b)$ & $B_b(M_Z)$ \\
\hline
$D^0$         & LO  & 72.8  & 67.6  & 57.5 & 52.7 \\
              & NLO & 71.6  & 65.8  & 54.3 & 49.3 \\
$D^+$         & LO  & 28.9  & 26.8  & 19.0 & 17.7 \\
              & NLO & 26.4  & 24.3  & 18.5 & 17.1 \\
$D^{*+}$      & LO  & 29.0  & 27.2  & 24.3 & 23.1 \\
              & NLO & 27.8  & 25.9  & 24.5 & 22.8 \\
$D_s^+$       & LO  & 12.3  & 11.7  & 23.1 & 21.2 \\
              & NLO & 10.6  & 10.0  & 22.1 & 20.2 \\
$\Lambda_c^+$ & LO  &  6.17 &  6.06 & 15.1 & 13.7 \\
              & NLO &  6.12 &  5.87 & 14.3 & 12.8 \\
\hline\hline
\end{tabular}
\end{center}
\end{table}

\begin{table}
\begin{center}
\caption{Average momentum fractions of $Q\to X_c$ for $Q=c,b$ and the various
$X_c$ hadrons evaluated according to Eq.~(\ref{eq:xq}) in LO and NLO at the
respective production thresholds $\mu=2m_Q$ and at the $Z$-boson resonance
$\mu=M_Z$.}
\label{tab:xav}
\begin{tabular}{cccccc}
\hline\hline
$X_c$ & Order & $\langle x\rangle_c(2m_c)$ & $\langle x\rangle_c(M_Z)$ &
$\langle x\rangle_b(2m_b)$ & $\langle x\rangle_b(M_Z)$ \\
\hline
$D^0$         & LO  & 0.573 & 0.442 & 0.318 & 0.285 \\
              & NLO & 0.550 & 0.420 & 0.304 & 0.272 \\
$D^+$         & LO  & 0.571 & 0.441 & 0.341 & 0.302 \\
              & NLO & 0.557 & 0.425 & 0.324 & 0.287 \\
$D^{*+}$      & LO  & 0.617 & 0.472 & 0.393 & 0.344 \\
              & NLO & 0.592 & 0.448 & 0.366 & 0.322 \\
$D_s^+$       & LO  & 0.654 & 0.496 & 0.348 & 0.310 \\
              & NLO & 0.653 & 0.487 & 0.337 & 0.299 \\
$\Lambda_c^+$ & LO  & 0.765 & 0.571 & 0.302 & 0.272 \\
              & NLO & 0.738 & 0.544 & 0.290 & 0.261 \\
\hline\hline
\end{tabular}
\end{center}
\end{table}

\begin{figure}[ht]
\begin{center}
\epsfig{file=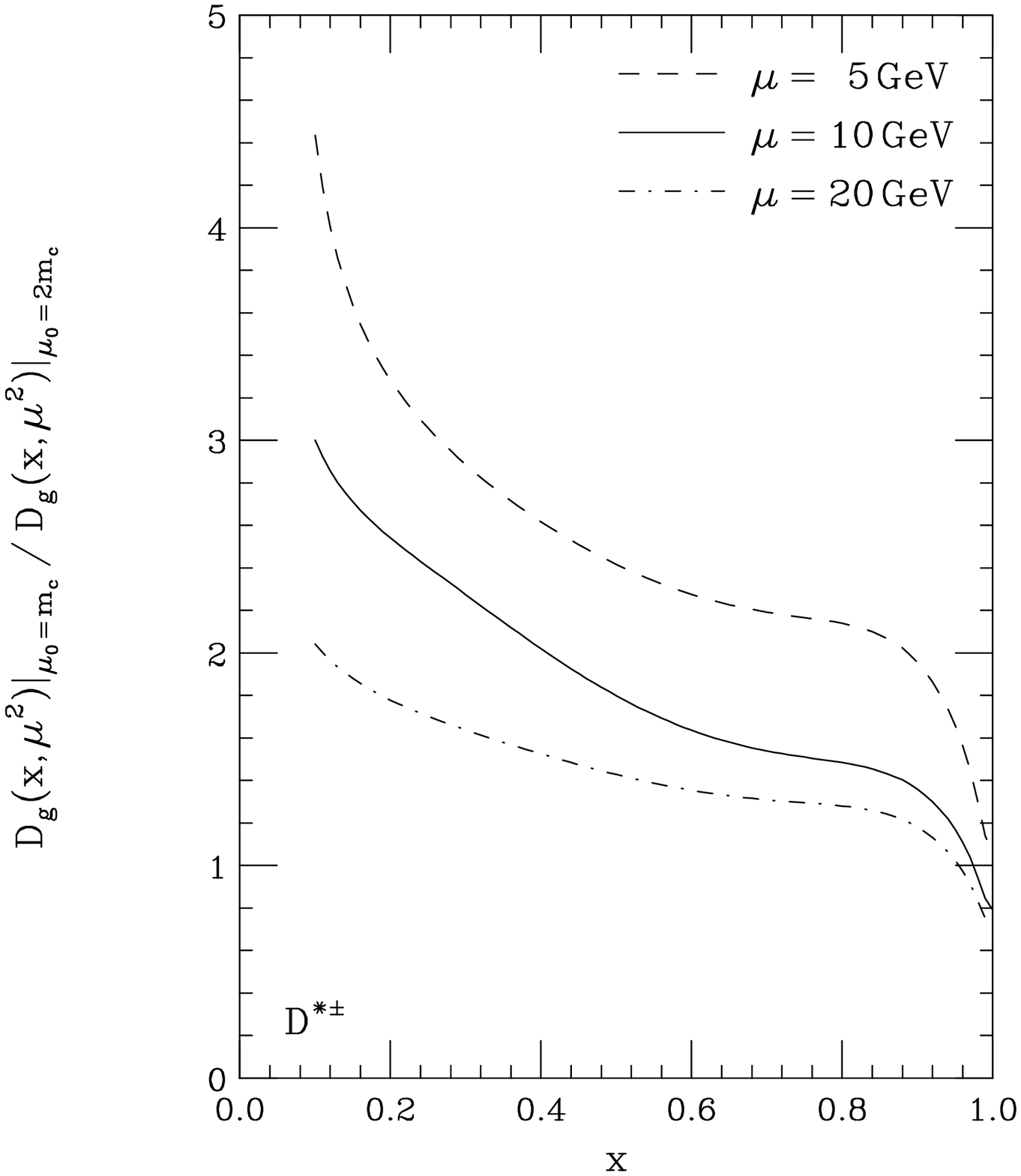,width=\textwidth}
\end{center}
\caption{$\mu_0=m_Q$ to $\mu_0=2m_Q$ ratios of $D_g(x,\mu^2)$ at $\mu=5$
(dashed), 10 (solid), and 20~GeV (dot-dashed) as functions of $x$ for
$X_c=D^{*+}$.}
\label{fig:one}
\end{figure}

For lack of space, we refrain from presenting here any updated versions of
figures included in Refs.~\cite{bkk,kk}; they would not exhibit any
qualitatively new features.
However, as already mentioned in Ref.~\cite{kkss}, the reduction in $\mu_0$
has an appreciable effect on the gluon FF's, which are only feebly constrained
by $e^+e^-$ data.
This effect is visualized for $X_c=D^{*+}$ in Fig.~\ref{fig:one}, where the
$\mu_0=m_Q$ to $\mu_0=2m_Q$ ratios of $D_g(x,\mu^2)$ at $\mu=5$, 10, and
20~GeV are shown as functions of $x$.
We observe that the reduction in $\mu_0$ leads to a significant enhancement of
the gluon FF, especially at low values of $x$.
The results for $X_c=D^0$, $D^+$, $D_s^+$, and $\Lambda_c^+$ are very similar
and, therefore, not shown here.

\section{Conclusions}

In this Addendum to Refs.~\cite{bkk,kk}, we repeated the fits of
non-perturbative $D^0$, $D^+$, $D^{*+}$, $D_s^+$, and $\Lambda_c^+$ FF's, both
at LO and NLO in the $\overline{\mathrm{MS}}$ factorization scheme, to OPAL
data from LEP1 \cite{opal,opal1} for the reduced choice $\mu_0=m_Q$ ($Q=c,b$)
of starting point for the DGLAP evolution in the factorization scale $\mu$.
These FF's are appropriate for use in connection with modern sets of PDF's
\cite{pdf}, which are implemented with the same convention for the
heavy-flavor thresholds.
A {\tt FORTRAN} routine that evaluates the values of these FF's as functions
of the input variables $x$ and $\mu$ may be obtained by electronic mail upon
request from the authors.

This reduction in $\mu_0$ is inconsequential for the theoretical
interpretation of experimental $e^+e^-$ data because it is compensated by
corresponding shifts in the fit parameters $N$, $\alpha$, $\beta$, and
$\epsilon$.
However, the gluon FF's, which are only feebly constrained by $e^+e^-$ data,
play a significant role in hadroproduction.
In fact, detailed analysis \cite{kkss} revealed that the increase in the gluon
FF's due to the extension of the evolution length leads to a rise in cross
section and thus improves the agreement with the CDF data of charmed-meson
production in run~II at the Tevatron \cite{cdf}.

\bigskip
\centerline{\bf ACKNOWLEDGMENTS}
\smallskip\noindent

We thank I.~Schienbein and H.~Spiesberger for useful discussions.
This work was supported in part by the Bundesministerium f\"ur Bildung und
Forschung through Grant No.\ 05~HT1GUA/4.


\begin{thebibliography}{99}

\bibitem{opal} OPAL Collaboration, K.~Ackerstaff {\it et al.},
Eur.\ Phys.\ J. C {\bf1}, 439 (1998).

\bibitem{opal1} OPAL Collaboration, G.~Alexander {\it et al.},
Z.\ Phys.\ C {\bf72}, 1 (1996).

\bibitem{bkk} J.~Binnewies, B.~A.~Kniehl, and G.~Kramer,
Phys.\ Rev.\ D {\bf58}, 014014 (1998).

\bibitem{kk} B.~A.~Kniehl and G.~Kramer,
Phys.\ Rev.\ D {\bf71}, 094013 (2005).

\bibitem{pet} C.~Peterson, D.~Schlatter, I.~Schmitt, and P.~M.~Zerwas,
Phys.\ Rev.\ D {\bf27}, 105 (1983).

\bibitem{col} J.~C.~Collins,
Phys.\ Rev.\ D {\bf58}, 094002 (1998).

\bibitem{dglap} V.~N.~Gribov and L.~N.~Lipatov,
Yad.\ Fiz.\ {\bf15}, 781 (1972)
[Sov.\ J. Nucl.\ Phys.\ {\bf15}, 438 (1972)];
G.~Altarelli and G.~Parisi,
Nucl.\ Phys.\ {\bf B126}, 298 (1977);
Yu.~L.~Dokshitzer,
Zh.\ Eksp.\ Teor.\ Fiz.\ {\bf73}, 1216 (1977)
[Sov.\ Phys.\ JETP {\bf46}, 641 (1977)].

\bibitem{rest} L.~Gladilin,
Report No.\ hep-ex/9912064 (unpublished);
DELPHI Collaboration, P.~Abreu {\it et al.},
Eur.\ Phys.\ J. C {\bf12}, 225 (2000);
ALEPH Collaboration, R.~Barate {\it et al.},
{\it ibid.}\ {\bf16}, 597 (2000);
S. Padhi, in {\it Proceedings of the Ringberg Workshop on New
Trends in HERA Physics 2003}, edited by G.~Grindhammer, B.~A.~Kniehl,
G.~Kramer, and W.~Ochs, (World Scientific, Singapore, 2004), p.~183;
H1 Collaboration, A.~Aktas {\it et al.},
Eur.\ Phys.\ J. C {\bf38}, 447 (2005).

\bibitem{cleo} CLEO Collaboration, D.~Bortoletto {\it et al.},
Phys.\ Rev.\ D {\bf37}, 1719 (1988);
CLEO Collaboration, R.~A.~Briere {\it et al.},
{\it ibid.}\ {\bf62}, 072003 (2000);
CLEO Collaboration, M.~Artuso {\it et al.},
{\it ibid.}\ {\bf70}, 112001 (2004).

\bibitem{hrs} HRS Collaboration, M.~Derrick {\it et al.},
Phys.\ Rev.\ Lett.\ {\bf54}, 2568 (1985);
HRS Collaboration, P.~Baringer {\it et al.},
Phys.\ Lett.\ B {\bf206}, 551 (1988).

\bibitem{tasso} TASSO Collaboration, M.~Althoff {\it et al.},
Phys.\ Lett.\ {\bf136B}, 130 (1984).

\bibitem{pdf} A.~D.~Martin, R.~G.~Roberts, W.~J.~Stirling, and R.~S.~Thorne,
Phys.\ Lett.\ B {\bf604}, 61 (2004);
J.~Pumplin, A.~Belyaev, J.~Huston, D.~Stump, and W.-K.~Tung,
JHEP {\bf0602}, 032 (2006).

\bibitem{kkss} B.~A.~Kniehl, G.~Kramer, I.~Schienbein, and H.~Spiesberger,
Phys.\ Rev.\ Lett.\ {\bf96}, 012001 (2006).

\bibitem{cdf} CDF Collaboration, D.~Acosta {\it et al.},
Phys.\ Rev.\ Lett.\ {\bf91}, 241804 (2003).

\bibitem{cac}  M.~Cacciari, P.~Nason, and C.~Oleari,
JHEP {\bf 0510}, 034 (2005).

\end{thebibliography}
\end{document}